\documentclass[aip,preprint]{revtex4-2}  

\usepackage{graphicx}
\usepackage{dcolumn}
\usepackage{bm}

\usepackage[utf8]{inputenc}
\usepackage[T1]{fontenc}
\usepackage{mathptmx}
\usepackage{etoolbox}
\usepackage{amssymb}
\usepackage[english]{babel}

\usepackage[english]{babel}

\usepackage[colorlinks=true,
            linkcolor=blue,
            citecolor=blue,
            urlcolor=blue]{hyperref}

\begin{document}


\title{Beyond the interface: Persistent Hopping Transport and Frequency Dispersion in Strong-inversion Cryogenic MOSFETs} 

\author{Keito Yoshinaga}
 \email{yoshinaga@ssn.t.u-tokyo.ac.jp}
\author{Wataru Miyagi}
\author{Ryo Toyoshima}
 \affiliation{Department of Materials Engineering, The University of Tokyo, Tokyo, Japan}
\author{Munehiro Tada}
 \affiliation{Faculty of Science and Engineering, Keio University, Yokohama, Japan}
\author{Ken Uchida}
\email{uchidak@material.t.u-tokyo.ac.jp}
 \affiliation{Department of Materials Engineering, The University of Tokyo, Tokyo, Japan}
\date{\today}

\begin{abstract}
  Cryogenic complementary metal-oxide-semiconductor (cryo-CMOS) technology is essential for quantum computing interfaces, which require
  precise modeling of dynamic device behavior. The output impedance of MOS field-effect transistors (MOSFETs) is frequency dependent, which
  has been conventionally attributed to extrinsic parasitics. Here, we report an intrinsic frequency dispersion in the channel impedance of
  cryogenic MOSFETs that persists deep into the strong-inversion region. Through a Cole–Cole analysis, we characterize this dispersion as a
  depressed semicircle in the impedance plane and attribute its behavior to variable-range hopping through band-tail localized states. Unlike
  conventional models where band-tail states are confined to the oxide interface, we demonstrate that in MOSFETs with high channel doping
  the band-tail states are induced by ionized impurities and distributed throughout the depletion region. Our paradigm accounts for frequency
  dispersion under strong inversion. This work demonstrates that ionized-impurities-induced hopping governs the dynamic response of
  cryo-MOSFETs channel impedance even when drift conduction dominates, offering critical insights for accurate small-signal modeling and
  high-frequency cryo-CMOS circuit design.
\end{abstract}


\maketitle 

Complementary metal-oxide-semiconductor technology operating at cryogenic temperatures (cryo-CMOS) has emerged as a key platform for quantum
readout and control circuits~\cite{Patra_2018}. To realize large-scale, high-fidelity quantum computers, the cryo-CMOS circuit design
requires an accurate prediction of dynamic behavior of CMOS devices at cryogenic temperatures. The small-signal output impedance of
MOS field-effect transistors (MOSFETs) is important, since it determines the frequency response, signal integrity, and noise performance of
cryo-CMOS devices. In conventional compact models, the channel (inversion layer) is treated as a quasi-static resistive
element~\cite{Incandela_2018,Singh_2022,Tada_2024}, and any frequency-dependent behavior is primarily attributed to external capacitances
and extrinsic parasitic effects.

In this study, through impedance measurements of cryogenic MOSFET channels, we observed a clear deviation from the conventional resistive
model, which is characterized by a capacitive response originating from variable-range hopping (VRH). The measured Cole–Cole (Nyquist)
plots exhibit a depressed semicircle in the impedance plane, clarifying intrinsic frequency dispersion and distributed charge-relaxation
dynamics. The frequency-dependent response persists deep into the strong-inversion regime, demonstrating that time-dependent relaxation
processes remain even when drift conduction in the inversion channel is dominant above the threshold voltage. By analyzing the
temperature-dependent subthreshold characteristics, we attribute the nonquasi-static behavior to VRH transport through band-tail localized
states. We found that, unlike the conventional paradigm, where band-tail states are confined to the oxide interface, these states are
distributed throughout the depletion region.  The potential fluctuations induced by ionized impurities generate these distributed states,
which account for the persistent frequency dispersion observed under strong inversion regime.  Finally, we validate the proposed physical
model through an equivalent circuit analysis of the gate-voltage dependence of the impedance.  We believe that the present findings are
useful for optimizing and designing future cryo-CMOS devices and circuits.

We prepared bulk Si $n$-type MOSFET with a substrate impurity concentration of $N_{\mathrm{sub}} = 1.3 \times 10^{17}~\mathrm{cm^{-3}}$,
which was determined from capacitance-voltage ($C-V$) measurements and was further confirmed with secondary ion mass spectrometry
analysis. The gate insulator was $\mathrm{SiO_2}$ with 50~nm thickness and the channel length ($L$) and width ($W$) were 100~$\mu\mathrm{m}$
and 100~$\mu\mathrm{m}$, respectively. For comparison, fully-depleted (FD) silicon-on-insulator (SOI) MOSFETs fabricated on lightly doped
substrates were also characterized. The nominal resistivity of the SOI layer was 10~$\Omega$cm, corresponding to an approximate impurity
concentration of $10^{15}~\mathrm{cm^{-3}}$. The gate oxide thickness, channel length and width were 10~nm, 100~$\mu\mathrm{m}$, and
100~$\mu\mathrm{m}$, respectively. The bulk and FD-SOI MOSFETs employed a common layout, with the primary structural difference being the gate
oxide thickness.

The interface state densities ($D_\mathrm{it}$) were measured using the charge pumping technique with the reported
methods~\cite{Uchida_2003,Li_1998,Ouisse_1991}, which resulted in $D_\mathrm{it}$ of $3\times10^{10}~\mathrm{cm}^{-2}\mathrm{eV^{-1}}$ and
$4\times10^{10}~\mathrm{cm}^{-2}\mathrm{eV^{-1}}$ for bulk and FD-SOI devices, respectively, showing the identical $D_\mathrm{it}$
levels. The channel impedance of the MOSFETs was measured using a Keysight E4980A LCR meter while a gate bias was applied by a Keithley
2636B source measure unit. The drain voltage for the impedance measurements consisted of a 5~$\mathrm{mV}_\mathrm{rms}$ AC signal
superimposed on a 5~mV DC bias. To eliminate parasitics, open-circuit corrections were performed under cutoff conditions. The DC electrical
characteristics were obtained using a Keysight 4156C semiconductor parameter analyzer with the drain voltage of 5~mV or lower to minimize
the pinch-off effects. All measurements were performed on a LakeShore CPX probe station. The stage temperature was controlled from 300 to
4.2~K.

\begin{figure*}
    \includegraphics[width=\textwidth]{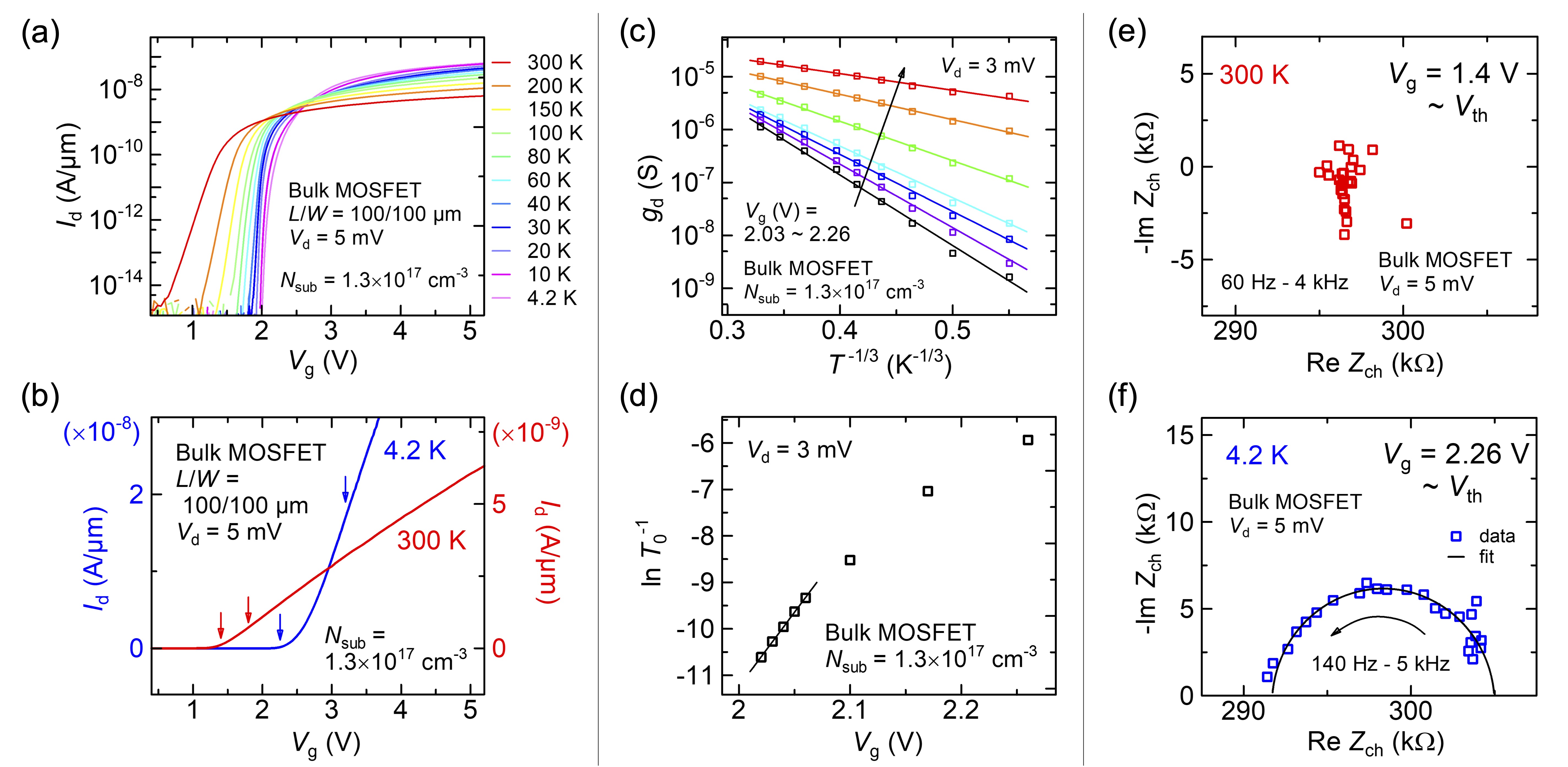}
    \caption{\label{subth} (a, b) Drain-current versus gate-voltage $I_\mathrm{d}-V_\mathrm{g}$ characteristics at various stage temperatures
      plotted on (a) logarithmic and (b) linear scales. The arrows in (b) indicate the gate voltages at which Cole–Cole plots are shown in
      Figs.~\ref{subth} and \ref{strong_inv}. (c) Temperature dependence of drain conductance $g_\mathrm{d}$ in the subthreshold regime,
      exhibiting a $T^{-1/3}$ dependence strongly indicative of hopping-dominated transport in the 2D electron system. (d) $V_\mathrm{g}$
      dependence of the hopping parameter $1/T_0$, which is proportional to the localized DOS at $E_\mathrm{F}$. $1/T_0$ exponentially
      increases with $V_\mathrm{g}$, reflecting the rapid increase in density of band-tail localized states at higher $V_\mathrm{g}$. (e, f)
      Cole–Cole plots of the channel impedance of bulk MOSFET at $V_\mathrm{g} \sim V_\mathrm{th}$. (e) At 300~K, the response is purely
      resistive; (f) at 4~K, a depressed semicircle characteristic of VRH conduction emerges.}
\end{figure*}

Figure~\ref{subth}a shows the drain current ($I_\mathrm{d}$) versus gate-voltage ($V_\mathrm{g}$) characteristics at various temperatures.
The same data for the temperatures of 300~K and 4.2~K are shown in Fig.~\ref{subth}b to illustrate the gate bias points for impedance
measurements. Figure~\ref{subth}c shows the drain conductance as a function of $T^{-1/3}$ for various $V_\mathrm{g}$ within the subthreshold
region, where $T$ is the temperature, indicating a clear linear trend in log-linear plot. This linear behavior confirms that $g_\mathrm{d}$
follows $g_\mathrm{d} \propto \exp \left[(T/T_\mathrm{0})^{-1/3}\right]$, which is the characteristic temperature dependence of
two-dimensional VRH transport~\cite{Mott_1969}, where $1/T_0$ is the hopping parameter for each $V_\mathrm{g}$. This result clearly
indicates that in the subthreshold region the Fermi level ($E_\mathrm{F}$) lies in the band-tail states and that the carrier transport is
dominated by hopping conduction between localized states. The hopping parameter $1/T_\mathrm{0}$ is proportional to the density of states
(DOS) of band-tail localized states near $E_\mathrm{F}$, which increases exponentially in the subthreshold region as a function of
$V_\mathrm{g}$, as shown in Fig.~\ref{subth}d. At higher $V_\mathrm{g}$, the exponential growth rate of the DOS moderates. These trends are
consistent with the fact that the band-tail DOS exponentially increases with energy~\cite{Zittartz_1966} and its growth rate subsequently
moderates toward the mobility edge~\cite{Thouless_1978}, demonstrating the validity of the fitting based on the VRH model. Thus, we
confirmed that in the subthreshold region of the cryo-CMOS transistors, carrier transport is dominated by VRH conduction.  Although several
recent studies on cryo-CMOS transistors~\cite{Kang_2023,Beckers_Inflection_2020} discuss the band-tail states, these reports consider that
one part of the band-tail states contribute to the drift-diffusion transport due to “mobile electrons” in these states and that the other
part is completely localized (immobile). However, hopping conduction does not require such a distinction between mobile and immobile states;
all localized states with energies below the mobility edge and around $E_\mathrm{F}$ naturally contribute to VRH conduction. Furthermore,
drift-diffusion and VRH transports have different characteristics including the temperature and bias dependencies.

Figures~\ref{subth}e and \ref{subth}f show the Cole–Cole plots of the channel impedance of bulk MOSFET ($V_\mathrm{g} \sim V_\mathrm{th}$) at
300~K and 4.2~K, respectively. The purely resistive response at 300~K transitions to a capacitive response at 4.2~K, under gate voltage
conditions where the impedance is comparable. The depressed semicircle in the Cole–Cole plot reflects an $RC$-like response with
distributed time constants~\cite{Cole_1941}, which appears only at low temperatures and is most plausibly explained by hopping
conduction~\cite{Miyao_2024}. This interpretation is validated by the temperature dependence of conductance at the same gate voltage
(2.26~V; Fig.~\ref{subth}c), which is another characteristic signal of hopping conduction discussed previously. Indeed, such
frequency-dependent impedance arising from hopping conduction has been documented in impurity
semiconductors~\cite{Pollak_1961,Golin_1963,Miyao_2024} and amorphous materials~\cite{Argall_1968,Ivkin_1970,Rockstad_1972,Frost_1975}.

\begin{figure*}
  \includegraphics[width=\textwidth]{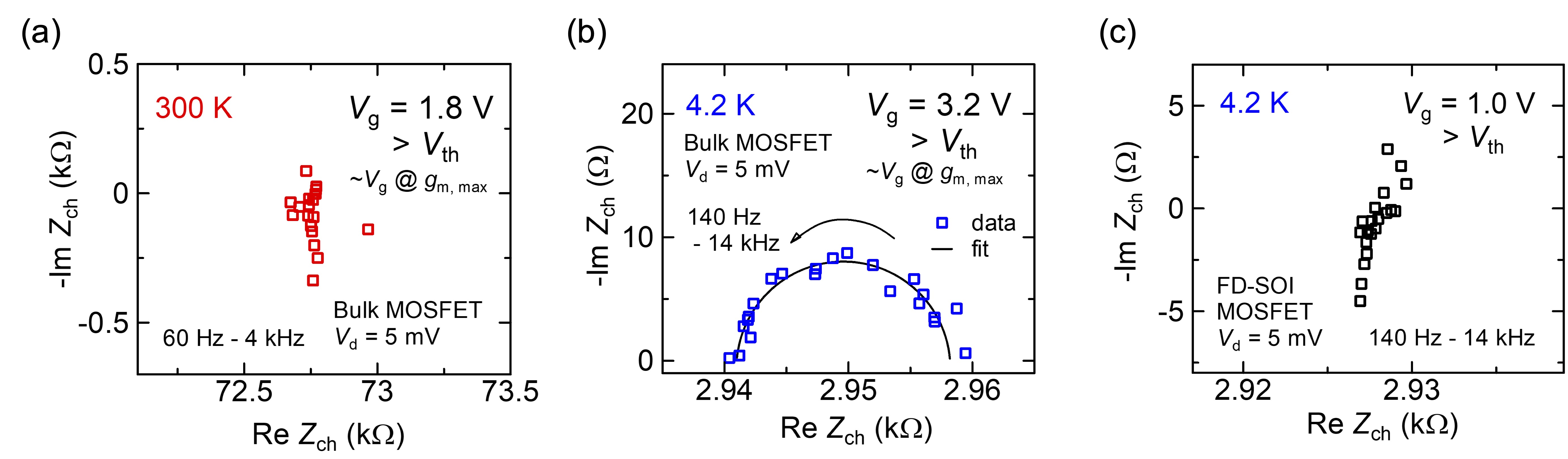}
    \caption{\label{strong_inv} Cole–Cole plots at $V_\mathrm{g} > V_\mathrm{th}$. (a) At 300~K, the response is purely resistive. (b) At
      4.2~K, the depressed semicircle indicates persistence of the VRH conduction path in the inversion regime, and (c) the channel
      impedance of the FD-SOI MOSFET exhibits no $RC$ response at 4.2~K.}
\end{figure*}

\begin{figure}
    \includegraphics[width=8.4cm]{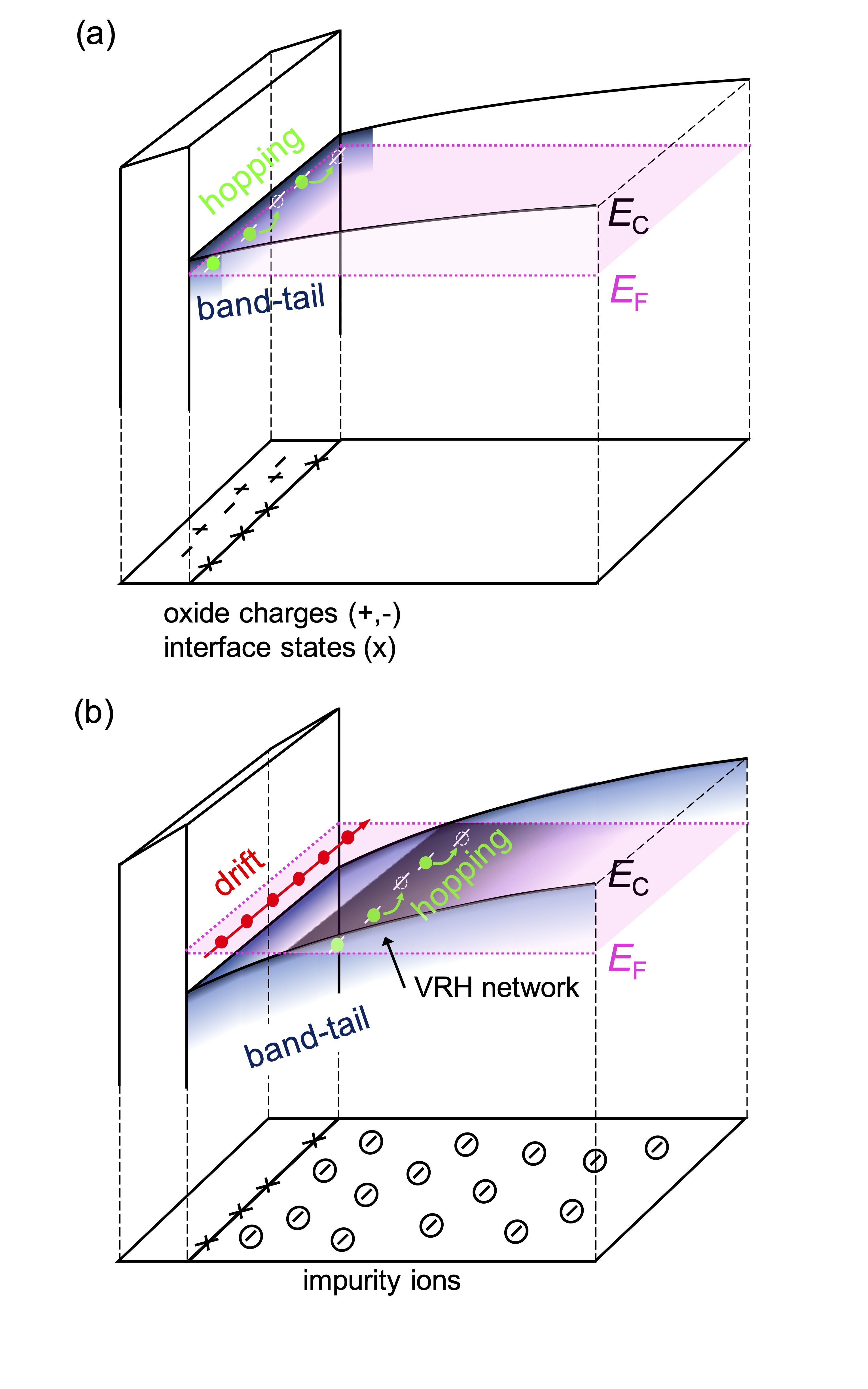}
    \caption{\label{VRH-network} (a) Conventional models assume that band-tail states are confined to the interface. (b) In
      reality, band-tail states are distributed throughout the depletion region hosting ionized dopants. Above $V_\mathrm{th}$, transport
      is dominated by drift conduction near the interface but parallel hopping conduction contributes to the AC response.}
\end{figure}

Figures~\ref{strong_inv}a and \ref{strong_inv}b show the Cole–Cole plots at 300~K and 4.2~K, respectively, measured at
$V_\mathrm{g} > V_\mathrm{th}$, specifically, at the gate voltage giving the maximum transconductance. At 300~K, the response is purely
resistive but at 4.2~K, the frequency dependence stemming from hopping conduction remains observable although drift conduction in the
inversion layer is expected to dominate the transport mechanism at voltages above the threshold voltage $V_\mathrm{th}$.  Although
cryo-MOSFETs have been characterized and modeled in recent studies, the physical origin of the band-tail states has been either largely
ignored or attributed to interface traps~\cite{Richstein_2022,Oka_noise_2023,Oka_2023} (Fig.~\ref{VRH-network}a). Observations of hopping
conduction in channels in the 1970s similarly focused exclusively on the band-tail states caused by oxide charges and interface roughness of
MNOS structures on Si (111) surfaces~\cite{Pepper_1974} or by intentionally introduced charges into the gate
oxide~\cite{Hartstein_1975}. However, in the present device such intentional charges were not introduced and low $D_\mathrm{it}$ was
confirmed. Therefore, we consider that localized band-tail states are generated by the potential formed by impurity ions within the
depletion region. Furthermore, it should be noted that the band-tail states exist not only at the interface but also throughout the
depletion region (Fig.~\ref{VRH-network}b). While carrier freeze-out generally neutralizes dopants in the neutral bulk at cryogenic
temperatures, the impurities within the depletion region are fully ionized due to the electric field. Furthermore, since there are no mobile
carriers to screen the Coulomb potential in the depletion region, the electrostatic disorder becomes exceptionally strong, leading to the
formation of these distributed band-tail states.  This spatial distribution consequently allows a robust VRH network to be sustained deeper
within the inversion layer even under strong-inversion conditions. In the subthreshold region, hopping conduction occurs in the band-tail
states near the interface and contributes to the AC response (Fig.~\ref{VRH-network}a). In the inversion region, DC transport is considered
to be dominated by drift conduction near the interface. However, the VRH network contributes to the AC response (Fig.~\ref{VRH-network}b),
which is pronounced at low-to-medium frequencies ($\sim$14~kHz) in Cole-Cole plots shown in Fig.~\ref{strong_inv}b.

The depressed semicircle is not observed in FD-SOI MOSFETs with low substrate doping ($N_\mathrm{sub}$) (Fig.~\ref{strong_inv}c). This is
reasonable, since potential fluctuations induced by ionized impurities are greatly suppressed in FD-SOI because of the two orders of
magnitude lower doping concentration in the channel. $D_\mathrm{it}$ of the FD-SOI MOSFET is the same as that of bulk MOSFET. Nevertheless, the
frequency dispersion is not observed in FD-SOI MOSFETs. Thus, the depressed semicircle, namely frequency dispersion, observed in bulk MOSFETs is
not due to the interface states but due to the potential fluctuations caused by ionized impurities in the depletion region.

\begin{figure}[htbp]
    \includegraphics[width=8.4cm]{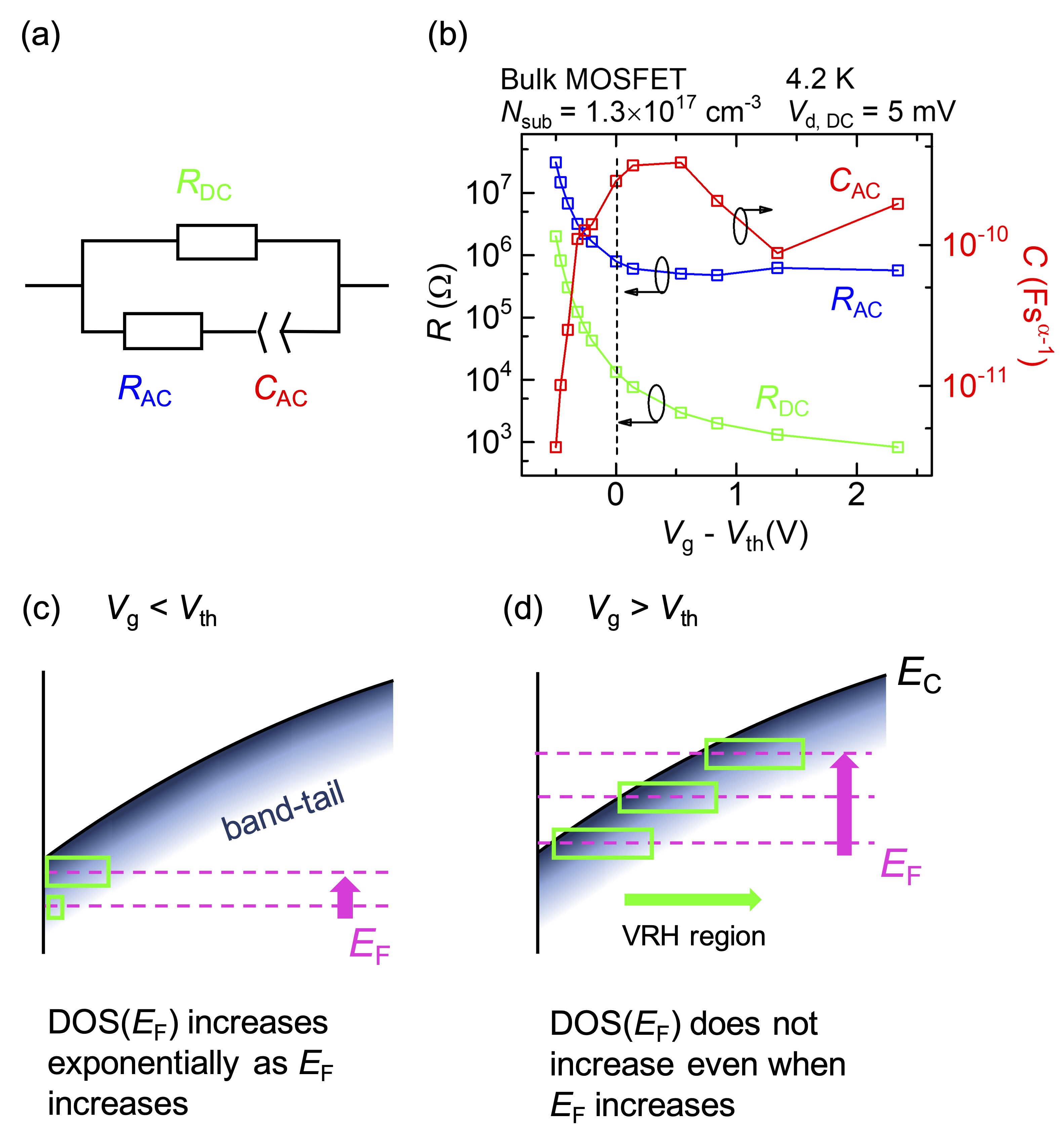}
    \caption{\label{circuit} (a) Equivalent circuit of the VRH network and inversion channel. (b) $V_\mathrm{g}$ dependence of the extracted
      circuit parameters. (c) In the subthreshold region $V_\mathrm{g} < V_\mathrm{th}$, hopping conductance exponentially increases with
      $V_\mathrm{g}$; (d) Above $V_\mathrm{th}$, the VRH network shifts deeper into the substrate with increasing $V_\mathrm{g}$ while the hopping
      conductance only modestly increases.}
\end{figure}

The impedance characteristics measured at each $V_\mathrm{g}$ were fitted to an equivalent circuit model shown in Fig.~\ref{circuit}a, which
consists of a DC resistance $R_{\mathrm{DC}}$ corresponding to the DC contributions of drift-diffusion and DC hopping conduction, a
resistance $R_{\mathrm{AC}}$ denoting the resistive component of AC hopping conduction, and a capacitance $C_{\mathrm{AC}}$ denoting the
capacitive component of AC hopping conduction. Figure~\ref{circuit}b shows the $V_\mathrm{g}$ dependence of the extracted
parameters. In the subthreshold region, the rapid decreases in $R_\mathrm{DC}$ and $R_\mathrm{AC}$ along with the sharp increase in
$C_\mathrm{AC}$ correspond to the increased subthreshold current, reflecting the exponential increase in the DOS that contributes to hopping
conduction (Fig.~\ref{circuit}c, see also Fig.~\ref{subth}d). Above the threshold voltage, $R_\mathrm{DC}$ decreases monotonically whereas
the $R_\mathrm{AC}$ and $C_\mathrm{AC}$ saturate. The monotonic decrease in $R_\mathrm{DC}$ reflects an increase in inversion-layer charge
density contributing to drift conduction, where $1/R_\mathrm{DC}$ approximately linearly depends on $V_\mathrm{g}$. Meanwhile, the
saturation of $R_\mathrm{AC}$ and $C_\mathrm{AC}$ above the threshold indicates a saturation of the DOS contributing to hopping conduction,
showing that as the gate voltage increases, the VRH network shifts toward the substrate side where the DOS at the Fermi level
($E_\mathrm{F}$) remains largely constant as shown in Fig.~\ref{circuit}d.

Understanding of the excess noise, subthreshold slope, and transient responses at cryogenic temperature is essential in cryo-CMOS
research. Although these phenomena have been discussed in several studies~\cite{Beckers_2020, Oka_noise_2023, Asanovski_2023, Cardoso_2020,
  Kiene_2024, Miyao_2022, Beckers_2025, Kang_2023, Oka_2023}, a unified understanding remains lacking. The present results show that the
ionized-impurity-induced band-tail states are distributed throughout the depletion region and play a critical role in the dynamic responses
of cryogenic MOSFETs. Our results can directly enhance the modeling and design of cryo-CMOS circuits, where distributed band-tail states may
contribute to excess noise. It is also clearly shown that hopping conduction dominates the subthreshold transport.  Therefore, it should be 
necessary to take into account the accurate modeling of band-tail localized DOS induced by ionized impurities~\cite{Yoshinaga_2025} for
the modeling of the hopping component in the drain currents.

Asanovski \textit{et al.} demonstrated that the cryogenic excess $1/f$ noise persists into the strong-inversion region, attributing this to
the kinetics of band-tail states~\cite{Asanovski_2023}. However, these states are generally considered, including their work, to be
localized in the immediate vicinity of the oxide interface~\cite{Pepper_1974, Oka_2023, Asanovski_2023}. Under strong
inversion, the $E_\mathrm{F}$ at the surface is pushed into the conduction band, leaving such interface states completely filled and
incapable of exchanging carriers. Incorporating the present paradigm offers one potential way to reconcile this contradiction, because the
ionized-impurity-induced band-tail states are distributed throughout the entire depletion region, indicating that at the other edge of the inversion
layer $E_\mathrm{F}$ lies in the band-tail states even under strong inversion condition as shown in Fig.~\ref{VRH-network}b. This
configuration allows these states to participate in the trap-detrap processes. Nevertheless, further systematic investigations are required
to fully clarify whether the primary origin of this excess noise is indeed the band-tail states within the depletion layer as we
propose, or if alternative physical mechanisms dominate.

In conclusion, this study demonstrated intrinsic frequency dispersion of the channel impedance of cryogenic MOSFETs, which persists into
strong inversion. This behavior originates from electrostatic disorder induced by ionized dopants in the depletion region, which produces
band-tail localized states. Hopping transport through these states manifests as a time-dependent response, demonstrating a non-quasi-static
component in MOSFET channel impedance both in subthreshold and inversion regions. Deviating from conventional models in which the band-tail states 
occurs near the oxide film or interface, our new physical paradigm accounts for band-tail states distributed throughout the entire depletion
layer. These findings can improve the accuracy of small-signal models and guide the design of cryo-CMOS circuits.

We thank Dr.~Michimasa Morita for valuable comments and discussions. This work was supported by JST Moonshot (JPMJMS2067 \& JPMJMS256F),
ASPIRE (JPMJAP2411), and KAKENHI (25H00731).

\section*{AUTHOR DECLARATIONS}
\subsection*{Conflict of Interest}
\noindent
The authors have no conflicts to disclose.

\subsection*{Author Contributions}
\noindent {\bf Keito Yoshinaga}: Data curation(lead); Formal analysis (lead); Methodology (equal); Writing -original draft (lead). {\bf
  Wataru Miyagi}: Data curation (equal); Formal analysis (supporting). {\bf Ryo Toyoshima}: Formal analysis (supporting); Writing -review \&
editing (supporting). {\bf Munehiro Tada}: Conceptualization (supporting); Funding acquisition (supporting); Writing -review \& editing
(supporting). {\bf Ken Uchida}: Conceptualization(lead); Formal analysis (equal); Methodology (lead); Data curation (equal); Writing -review
\& editing (lead); Funding acquisition (lead); Resources (lead).

\section*{DATA AVAILABILITY}
\noindent
The data supporting the findings of this study are available from the corresponding author upon reasonable request.

\bibliography{VRH_CryoCMOS}

\end{document}